\documentstyle[11pt]{article}

\def\be{\begin{equation}}
\def\ee{\end{equation}}
\def\bea{\begin{eqnarray}}
\def\eea{\end{eqnarray}}

\def\l{\label}
\def\c{\cite}
\def\r{\ref}
\def\fn{\footnote}
\renewcommand{\theequation}{\arabic{section}.\arabic{equation}}

\def\al{\alpha}

\def\b{\beta}

\def\epsi{\epsilon_1}
\def\epsii{\epsilon_2}

\def\be{\begin{equation}}
\def\ee{\end{equation}}
\def\bea{\begin{eqnarray}}
\def\eea{\end{eqnarray}}

\def\nn{\nonumber}

\def\wt{\widetilde}
\def\nab{\nabla} 
\def\case#1/#2{\textstyle\frac{#1}{#2}}
\def\k0{\kappa_{0}}

\def\fr{\frac}
\def\L{\left}
\def\R{\right}

\begin{document}
\begin{titlepage}

\vspace{.7in}

\begin{center}
\Large
{\bf Inhomogeneous Einstein--Rosen String Cosmology}\\
\vspace{.7in}
\normalsize
\large{Dominic Clancy$^{1a}$, 
Alexander Feinstein$^{2b}$, James E. Lidsey$^{3c}$
\& Reza Tavakol$^{1d}$}\\
\normalsize
\vspace{.4in}
$^1${\em Astronomy Unit, School of Mathematical Sciences,  \\
Queen Mary \& Westfield College, Mile End Road, London, E1 4NS, U.K.}\\
\vspace{.2in}
$^2${\em Dpto. de F\'{\i}sica Te\'orica, Universidad del Pa\'{\i}s Vasco, \\
Apdo. 644, E-48080, Bilbao, Spain}\\
\vspace{.2in}
$^3${\em Astronomy Centre and Centre for Theoretical Physics, \\
University of Sussex, Brighton, BN1 9QH, U.K.}
\end{center}
\vspace{.3in}
\baselineskip=24pt

\begin{abstract}
\noindent Families of anisotropic and 
inhomogeneous string cosmologies containing 
non--trivial dilaton and axion fields are derived 
by applying the global symmetries of the string effective action
to a generalized Einstein--Rosen metric. The models exhibit a 
two--dimensional group of Abelian isometries. In particular, 
two classes of exact solutions are found that represent 
inhomogeneous generalizations of the Bianchi type ${\rm VI}_h$ 
cosmology. The asymptotic behaviour of the solutions is 
investigated and further applications are briefly discussed. 

\end{abstract}

PACS NUMBERS:  04.20.Jb, 04.50.+h, 11.25.Mj, 98.80.Cq   

\vspace{.3in}
$^a$Electronic address: dominic@maths.qmw.ac.uk

$^b$Electronic address: wtpfexxa@lg.ehu.es

$^c$Electronic address: jel@astr.cpes.susx.ac.uk

$^d$Electronic address: reza@maths.qmw.ac.uk

\end{titlepage}

\section{Introduction}

\setcounter{equation}{0} 
\def\theequation{\thesection.\arabic{equation}}

An important lesson of nonlinear dynamical systems theory is that
solutions originating from different regions of state space
can (and often do) have qualitatively different modes
of behaviour. In this sense,  the initial conditions
become an important ingredient in determining the
dynamical evolution of the system. This then raises the question of
how one determines the set of initial conditions that gives rise to a
particular dynamical mode of behaviour. The answer to this question
requires a full understanding  of the underlying dynamics which,  in the 
case of general relativity or string theory, 
is extremely difficult to establish,
given the high degree of nonlinearity and complexity
involved in these theories. There is also the added difficulty that
despite recent progress
in our understanding of M--theory, a definitive non--perturbative
formulation of quantum gravity still remains to be developed 
(see, e.g., Ref. \cite{Mth} for recent reviews).
 
This question is particularly relevant in cosmological 
studies of string or M--theory. The low-energy effective action of the 
Neveu--Schwarz/Neveu--Schwarz (NS--NS) bosonic sector of string theory 
contains a multiplet of massless fields 
$\{ g_{\mu\nu}, \phi ,B_{\mu\nu}\}$ \cite{GSW}. 
The vacuum expectation value of the
dilaton,  $\phi$,  determines the string (gravitational) 
coupling, $g_s^2 \equiv e^{\phi}$, 
the gravitational field is determined by the metric,  
$g_{\mu\nu}$, and $B_{\mu\nu}$ 
is an antisymmetric, two--form potential. 
To date, attention has focused on the spatially homogeneous, 
orthogonal models, where the dilaton field is constant on 
the surfaces of homogeneity \cite{hmod,clw,hmod1}. 
However, these models apply on scales 
just below the string scale, and it is precisely this region 
where spatial inhomogeneities may be important. 

A study of inhomogeneous  
string cosmologies is therefore necessary if further 
progress is to be made in addressing the question of whether our 
universe arose out of generic initial conditions. 
The purpose of the present paper is to derive and study 
a wide class of `Einstein--Rosen' string cosmologies with non--trivial 
dilaton and two--form potential. Einstein--Rosen metrics 
are interesting for a number of reasons \cite{einsteinrosen}.
(For reviews see, e.g., \cite{G2review1,G2review2}).  
Spatial homogeneity is broken along one direction and they 
admit an Abelian isometry group, $G_2$, 
that acts on two--dimensional spacelike orbits. They represent  
a natural generalization of the spatially 
homogeneous Bianchi cosmologies \cite{tomita,rs}. Density perturbations 
in the early universe can be analysed with these backgrounds 
and the propagation and collision of 
gravitational plane waves on homogeneous space--times 
can also be studied within this context 
\cite{collide,Adams1,Adams2,alex}. Finally, 
it has been conjectured that $G_2$ metrics represent a leading--order 
approximation to more general solutions near the 
singularity \cite{leading}.  

There has also been an interesting recent
development within string cosmology,
namely the {\em pre--big bang} scenario \cite{pbb},
according to which the rapid increase of the string coupling
drives an accelerated, inflationary expansion.
The central postulate of this scenario is that
the initial state of the universe is in the perturbative regime
of small coupling and curvature. This leads to an inflationary phase
for sufficiently homogeneous initial conditions \cite{qh,qh1}. At present,
the question of whether in general large spatial inhomogeneities
have a significant effect on the naturalness of such initial data
is unresolved.

Recently, Barrow and Kunze \cite{bk} studied 
a class of inhomogeneous generalizations of 
Bianchi I string cosmologies and Feinstein, Lazkoz and 
Vazquez--Mozo~\cite{flv} 
derived an inhomogeneous model by applying duality 
transformations on the locally rotationally symmetric (LRS) 
Bianchi type IX cosmology. 
In this paper we consider the 
$G_2$ inhomogeneous generalizations 
of the Bianchi type ${\rm VI}_h$ universe, 
where $h<0$ is the group parameter. This Bianchi model 
is interesting because it has a non--zero measure in the space 
of homogeneous initial data and includes the 
Bianchi type III as a special case $(h=-1)$ \cite{measure}. Furthermore, 
the most general spatially homogeneous 
solutions of the (one--loop) string equations of motion 
are the Bianchi types III and ${\rm VI}_h$, where $h=\{ 0, -1/2 , -2 \}$ 
\cite{bk}. 
It can be shown that 
these models contain the maximum number of eight free parameters.  

We employ 
non--compact, global symmetries of the field equations
to generate inhomogeneous 
solutions with a non--trivial two--form potential. 
When the metric admits two commuting spacelike Killing vectors, 
there exists an infinite--dimensional symmetry on the space of solutions
that may be identified infinitesimally 
with the ${\rm O}(2,2)$ current algebra \cite{bakas,bakas1}. This  
symmetry reduces to the Geroch group, 
corresponding to the ${\rm SL}(2,R)$ current 
algebra,  when the dilaton and two--form potential 
are trivial \cite{geroch}. The global ${\rm SL}(2,R)$ `S--duality' 
\cite{sref} and 
${\rm O}(2,2;R)$ `T--duality' \cite{tref} 
are contained within this symmetry\footnote{We 
refer to the groups ${\rm SL}(2,R)$ and ${\rm O}(2,2;R)$ 
as the  S-- and T--duality groups, respectively, although 
at the non--perturbative level the dualities are 
the discrete subgroups ${\rm SL}(2,Z)$ and ${\rm O}(2,2;Z)$.}.
Application of both these symmetries leads to new, inequivalent 
solutions. 

This paper is organised as follows. In Section 2, we
derive inhomogeneous $G_2$ string cosmologies 
from a general class of Einstein--Rosen models
where the two--form potential is trivial. 
Two families of solutions representing inhomogeneous generalizations of 
the Bianchi type ${\rm VI}_h$ universe  
are found in Section 3 by directly  solving the field equations. The 
asymptotic behaviour of such models is studied in Section 4. 
We conclude with a discussion in Section 5. 

\section{Einstein--Rosen String Cosmology}

\setcounter{equation}{0} 
\def\theequation{\thesection.\arabic{equation}}

\subsection{String Effective Action}

Fundamental strings sweep geodesic surfaces with respect 
to the string--frame metric, $g_{\mu\nu}$. The 
four--dimensional, string effective action for the NS--NS 
fields is given by 
\be
\l{effectiveaction}
S= \int d^4 x \sqrt{-g} e^{-\phi}
\left[  R +({\nab}\phi)^2-\fr{1}{12}H_{\mu\nu\lambda}H^{\mu\nu\lambda}
\right] ,
\ee
where $H_{\mu\nu\lambda} \equiv \partial_{[\mu}B_{\nu\lambda ]}$
is the field strength of the two--form potential. In general, 
the effective action will also include moduli 
and vector fields 
arising from the compactification from higher dimensions. The action  
is also expected to include a potential term, $V(\phi)$, arising 
from the non-perturbative sector of the theory, however, the form of 
such a potential is as yet unknown. These additional terms 
are neglected in what follows. 

In order to take advantage of the highly developed framework of general 
relativity and its many known exact solutions, it is often more convenient to 
work in the Einstein frame, where the dilaton field is minimally coupled to gravity. This is achieved by making the conformal 
transformation
\be
\wt{g}_{\mu\nu}=e^{-\phi}g_{\mu\nu}.
\ee
Action (\ref{effectiveaction}) then takes the form
\be
S= \int{d^4x\sqrt{-\wt{g}}
\L[\wt{R}-\frac{1}{2} 
(\tilde{\nabla} \phi)^2-\fr{1}{12}e^{-2\phi} 
\tilde{H}_{\mu\nu\lambda} \tilde{H}^{\mu\nu\lambda} \R]}.
\ee 
In four dimensions the field strength of the two--form 
potential is dual to a one--form: 
\be
\tilde{H}^{\mu\nu\lambda} = \tilde{\epsilon}^{\mu\nu\lambda\kappa} 
e^{2\phi} \tilde{\nabla}_{\kappa} \sigma  ,
\ee
where $\tilde{\epsilon}^{\mu\nu\lambda\kappa}$ is the covariantly
constant four--form \cite{sref}. The field equations can then 
be derived from the dual action
\be
\label{sigmaaction}
S=\int d^4 x \sqrt{-\tilde{g}}\left[ \tilde{R} -\frac{1}{2} 
\left( \tilde{\nabla} \phi \right)^2 - \frac{1}{2} e^{2\phi} \left(
\tilde{\nabla} \sigma \right)^2 \right]  ,
\ee
where $\sigma$ may be interpreted as a pseudo--scalar `axion' field. 

The generalized Einstein--Rosen $G_2$ metric  
is defined in the Einstein frame by the line element 
\cite{einsteinrosen,G2review1}
\be
\label{g2line}
ds^2_e=\tilde{h}_{\alpha\beta} (x^{\gamma}) dx^{\alpha} dx^{\beta}
+\tilde{\gamma}_{ab} (x^{\gamma}) dx^a dx^b,
\ee
where all components are taken to be 
independent of the spatial coordinates $x^a =(x, y)$.
The two commuting, spacelike Killing vectors are $\partial /\partial x$ 
and $\partial / \partial y$ and 
$\tilde{h}_{\alpha\beta}$ represents 
the longitudinal component of the metric. The metric on  the surfaces of 
transitivity is denoted by $\tilde{\gamma}_{ab}$ and 
the gradient $K_{\mu}
\equiv \partial_{\mu} \left( {\rm det} \tilde{\gamma}_{ab}
\right)^{1/2}$ determines the local behaviour of the model. 
Solutions represent cylindrical and plane gravitational 
waves if $K_{\mu}$ is globally spacelike or null, respectively 
\cite{G2review1,collide}. 
Cosmological models arise when $K_{\mu}$ is timelike or when 
the sign of $K_{\mu}K^{\mu}$ changes \cite{tomita}. 

The longitudinal metric is conformally flat and
the line element may therefore be written in the form 
\be
ds^2_e =e^f \L( -d\xi^2 +dz^2 \R)+\tilde{\gamma}_{ab}
dx^a dx^b ,
\ee
where  $f=f(\xi , z)$ determines the longitudinal part of the 
gravitational field. The corresponding line element in the string 
frame is given by
\begin{equation}
\label{stringgeneral}
ds_s^2 =e^{\phi + f} \left( -d\xi^2 +dz^2 \right) +
\Gamma_{ab} dx^adx^b   ,
\ee
where the transverse metric 
$\Gamma_{ab} \equiv e^{\phi} \tilde{\gamma}_{ab}$ 
has determinant $\Gamma \equiv {\rm det}\Gamma_{ab}$. 
We assume throughout this work that all massless fields are independent 
of the coordinates $x^a$. Thus, the metric (\ref{stringgeneral}) 
also represents a $G_2$ model. 

A  considerable 
simplification occurs in the field equations 
when the transverse metric is diagonal 
and separable. In this case, the Einstein and 
string frame metrics may be written in the form
\be
\l{canmetric}
ds^2_e=e^f(-d\xi^2+dz^2)+\xi(e^p dx^2+e^{-p} dy^2) 
\ee
and 
\be
\label{canmetricstring}
ds^2_s = e^{f+\phi} \left( -d\xi^2 +dz^2 \right) 
+\xi e^{\phi} \left( e^p dx^2 +e^{-p}dy^2 \right)   ,
\ee
respectively, where $p=p(\xi , z)$
represents  the transverse part of the gravitational field. 
In this case, the volume of the transverse space in the string frame 
is determined by 
\begin{equation}
\label{detvol}
\Gamma = \xi^2 e^{2\phi}   .
\end{equation}
In some settings, it proves convenient to define new variables
\bea
\label{emcoords}
z \equiv  e^{-2Z}\cosh(2t) \nonumber \\
\xi \equiv  e^{-2Z}\sinh(2t)   ,
\eea
which transform the metric (\ref{canmetric}) to 
\be
\label{tzmetric}
ds^2_e = 4 e^{h}(-dt^2 +dZ^2) + e^{-2Z}\sinh (2t)
(e^p dx^2+e^{-p} dy^2),
\ee
where $h \equiv f-4Z$.

In the present context, an inhomogeneous string 
cosmology is parametrized by the massless degrees of freedom 
$\{ g_{\mu\nu}, \phi, B_{\mu\nu} \}$. A vacuum 
solution to Einstein gravity may then 
be represented by $\{ g_{\mu\nu} , 0, 0\}$
and a solution with a trivial two--form potential by $\{ g_{\mu\nu}, 
\phi ,0\}$. We refer to this latter class of solution as `dilaton--vacuum' 
solutions. In general, the one--loop  field equations of motion for $G_2$ 
backgrounds derived from the actions (\ref{effectiveaction}) 
and (\ref{sigmaaction}) 
are difficult to solve. In view of this, we employ the non--compact, 
global symmetries that arise when the metric admits two 
Abelian isometries to generate a wide class of 
inhomogeneous string cosmologies with a non--trivial 
two--form potential from dilaton--vacuum solutions.

\subsection{O(2,2) Symmetry}

The global ${\rm O}(2,2)$ symmetry applies when
there exist two Abelian isometries and the 
only non--trivial component of the two--form potential 
is $B_{xy} = B_{xy}(\xi ,z)$ \cite{tref}. 
This symmetry is manifest in the string frame and 
generates fractional linear transformations on the two--form 
potential and the components of the transverse metric, $\Gamma_{ab}$. 
When the transverse metric is non--diagonal, the 
four degrees of freedom $\{ B_{xy} , \Gamma_{ab} \}$ 
parametrize the ${\rm O}(2,2)/[{\rm O}(2) \times {\rm O}(2)]$ coset
\cite{tref}. 
The isomorphism ${\rm O}(2,2) = {\rm SL}(2,R) \times {\rm SL}(2,R)$ 
then implies that these may be arranged in terms of two complex coordinates
\cite{Dij}
\bea
\tau \equiv \frac{\Gamma_{xy}}{\Gamma_{yy}}
+i\frac{\sqrt{\Gamma}}{\Gamma_{yy}} \\
\rho \equiv B_{xy} +i \sqrt{\Gamma}  .
\eea

Suppose the metric (\ref{canmetricstring}) represents a 
vacuum or dilaton--vacuum solution 
for some appropriate form of the dilaton, $\phi =\phi (\xi ,z)$. 
An ${\rm O}(2,2)$ transformation is then generated in terms of the 
two ${\rm SL}(2,R)$ transformations: 
\bea
\label{trho}
\bar{\rho} &=& \frac{a\rho + b}{c \rho +d} , \qquad \bar{\tau} =\tau \\
\label{ttau}
\bar{\tau} &=& \frac{a' \tau + b'}{c' \tau +d'} , \qquad \bar{\rho} 
= \rho ,
\eea
where $ad -bc =1$ and $a'd' - b'c' =1$. 
Under a general ${\rm O}(2,2)$ transformation, the dilaton transforms to 
\be
\label{tdilaton}
e^{\bar{\phi}} = e^{\phi} \left( \frac{\bar{\Gamma}}{\Gamma} 
\right)^{1/2}
\ee
and the longitudinal part of the string frame metric remains 
invariant, i.e., 
\be
\label{tlong}
\bar{f} = f +\phi -\bar{\phi} .
\ee

Since the transformations 
(\ref{ttau}) leave the two--form potential invariant, 
(\ref{trho}) must be employed to 
generate such a field from a dilaton--vacuum seed 
solution. Applying (\ref{trho}) implies that 
\bea
\label{tGamma}
\bar{\Gamma} &=& \frac{\Gamma}{\left( d^2 +c^2 \Gamma 
\right)^2}, \\
\label{taxion}
\bar{B}_{xy} &=& \frac{ac \Gamma +bd}{c^2 \Gamma +d^2}, \\
\label{tphi}
e^{\bar{\phi}} &=& \frac{e^{\phi}}{d^2 +c^2 \Gamma}, 
\eea
where the last transformation follows from (\r{detvol}) and (\r{tGamma}). 
The dual metrics in the string and Einstein frames are then given by 
\be
\label{tstringmetric}
d\bar{s}^2_s =e^{f +\phi} \left( -d\xi^2 + dz^2 \right) 
+\frac{\xi e^{\phi}}{d^2 +c^2 \xi^2 e^{2\phi}} 
\left( e^p dx^2 +e^{-p}dy^2 \right) 
\ee
and 
\be
\label{teinsteinmetric}
d\bar{s}_e^2 = e^f \left( d^2 +c^2 \xi^2 e^{2\phi} 
\right)  \left( -d\xi^2 + dz^2 \right) +\xi 
\left( e^p dx^2 +e^{-p}dy^2 \right)   ,
\ee
respectively. 
We remark that the transverse metric in the Einstein  frame, 
$\tilde{\gamma}_{ab}$,  
is invariant under the transformations (\ref{trho}). When $d=0$ and 
$c=1$, the volume of the transverse space in the string frame, 
as given by Eq. (\ref{tGamma}),  is 
inverted. The transformations (\ref{trho}) therefore represent a `T--duality'.

\subsection{SL(2,R) Symmetry}

The global ${\rm SL}(2,R)$ 
symmetry of the string effective action (\ref{effectiveaction})
becomes manifest in the 
Einstein frame. The action (\ref{sigmaaction}) may be written as a 
non--linear sigma--model, where  
the dilaton and axion fields parametrize the ${\rm SL}(2,R)/{\rm U}(1)$
coset \cite{sref}. The effective action 
is therefore invariant under global ${\rm SL}(2,R)$ transformations. 
These act non--linearly on the complex scalar field 
$\chi \equiv \sigma +ie^{-\phi}$ such 
that the transformed field is given by 
\begin{equation}
\label{chi}
\bar{\chi}
= \frac{A \chi +B}{C \chi +D}  ,
\end{equation}
where $\{A,B,C,D \}$ are real numbers satisfying 
$AD-BC =1$. The Einstein frame metric transforms as a singlet 
under this ${\rm SL}(2, R)$ transformation and the 
dual string frame metric is therefore given by 
\be
\label{sdualmetric}
d\bar{s}_s^2 =e^{\bar{\phi}-\phi} ds^2_s   .
\ee
For the special case where $C^2 =1$ and $\sigma = -D/C$, 
the ${\rm SL}(2,R)$ transformation (\r{chi}) yields 
$\bar{\phi} =-\phi$, which corresponds to an 
inversion of the string coupling $\bar{g}_s = g_s^{-1}$. 
The transformations (\ref{chi}) therefore 
represent a strong/weak--coupling `S--duality'. 

Starting with $\sigma=0$, a solution with non--trivial axion field may 
be generated directly from a given 
dilaton--vacuum solution of the generic form 
(\ref{canmetric}) by applying Eq. (\ref{chi}). The dual solutions 
are given by 
\be
\label{ds2-strings}
d\bar{s}^2_s = e^{\bar{\phi}+f} \left( 
-d\xi^2 +dz^2 \right) +\xi e^{\bar{\phi}} \left( 
e^p dx^2 +e^{-p} dy^2 \right) \\
\ee
\bea
\label{sphivacuum}
e^{\bar{\phi}} &=& C^2 e^{-\phi} +D^2e^{\phi} \\
\label{ssigmavacuum}
\bar{\sigma} &=& \frac{AC e^{-\phi} 
+BDe^{\phi}}{C^2e^{-\phi} + D^2e^{\phi}}   .
\eea

To summarize thus far,  we have seen how two 
inequivalent classes of inhomogeneous 
string cosmologies with a non--trivial two--form potential 
can be derived from a given dilaton--vacuum solution, 
by employing the non--compact, global ${\rm SL}(2,R)$ 
and ${\rm O}(2,2)$ symmetries of the model.
The dual solutions may be referred 
to as `dilaton--axion' cosmologies. In all 
cases, they are parametrized in terms 
of the functions $\{ f, p,  \phi \}$ that define the seed 
dilaton--vacuum solutions. 

Thus, the asymptotic behaviour of these models can be investigated 
directly once the seed solution has been specified. We therefore 
proceed in the following Section  to derive two classes of 
inhomogeneous dilaton--vacuum cosmologies that may be viewed 
as generalizations of the homogeneous Bianchi type ${\rm VI}_h$ 
universe. 

\section{Inhomogeneous Dilaton--Vacuum Cosmology}

\setcounter{equation}{0} 
\def\theequation{\thesection.\arabic{equation}}

\subsection{Cosmological Field Equations}

When the two--form potential vanishes, action (\ref{sigmaaction}) 
reduces to that for a massless,  minimally coupled scalar field. 
For the metric (\ref{canmetric}) the field equations then take the form 
\cite{cm}
\be
\l{efe1}
\dot{f}=-\fr{1}{2\xi}+ \frac{\xi}{2} \left( 
\dot{p}^2 + {p'}^2 +\dot{\phi}^2 +{\phi'}^2 \right)
\ee
\be
\l{efe2}
f'=\xi\left(\dot{p}{p'}+ \dot{\phi}{\phi'}\right),
\ee
\be
\l{efe3}
\ddot{p}+\fr{1}{\xi}\dot{p}-p''=0,
\ee
\be
\l{efe4}
\ddot{\phi}+\fr{1}{\xi}\dot{\phi}-\phi''=0,
\ee
in which overdots (primes) denote differentiation with respect to the
timelike variable $\xi$ (spacelike variable $z$). Eqs. (\r{efe3}) and
(\r{efe4}) are the integrability conditions for the system (\r{efe1}) 
and (\r{efe2}). 
The field equations (\ref{efe1})--(\ref{efe4}) are 
invariant under the simultaneous interchange $p 
\leftrightarrow \phi$. Indeed, the wave equations (\r{efe3}) and
(\ref{efe4}) are formally equivalent to the cylindrically symmetric 
wave equation in flat space, for which the general 
solution is known. A number of cosmologically relevant 
solutions satisfying different boundary conditions have been 
considered previously \cite{G2review1,G2review2,Gowdy}. 
An important feature of these equations is that
they are linear, which implies that new solutions may be constructed 
from superpositions of known solutions. 

Here we consider a linear 
superposition of solutions of the form: 
\be
\l{psoln}
p=k\ln(\xi)-m\cosh^{-1}\L(\fr{z}{\xi}\R)
+\epsi\int_0^\infty\left[c_1(z)J_0(l\xi)+c_2(z)N_0(l\xi)\right]\,dl
\ee
\be
\l{phisoln}
\phi = \alpha \ln(\xi)
- \beta  \cosh^{-1}\L(\fr{z}{\xi}\R)
+\epsii\int_0^\infty\left[c_3(z)J_0(l\xi)+c_4(z)N_0(l\xi)\right]\,dl   ,
\ee
where $\{ \alpha , \beta , k , m \}$ 
and $\{ \epsilon_1 , \epsilon_2 \}$ are constants, $J_0$ and $N_0$ are 
zero--order Bessel functions of the first and second kind 
and the coefficients $c_i=c_i(z)$ are defined by 
\bea
\l{c1}
c_1(z)&=&C_l\cos(lz)+D_l\sin(lz)  \\
\l{c2}
c_2(z)&=&F_l\cos(lz)+G_l\sin(lz)  \\
\l{c3}
c_3(z)&=&H_l\cos(lz)+L_l\sin(lz)  \\
\l{c4}
c_4(z)&=&U_l\cos(lz)+V_l\sin(lz),
\eea
where $C_l,\,D_l$, etc.,  are arbitrary constants for each 
value of $l$. In both Eqs. (\r{psoln}) 
and (\r{phisoln}), the last term represents 
the most general {\em separable} solution to Eqs. (\ref{efe3}) 
and (\ref{efe4}). 
The remaining two terms represent other, in general inhomogeneous, 
solutions\footnote{The  
first term is also present in the third term and
is therefore not a different solution. It is separated out for later 
consideration.}. The importance of these other solutions is that 
they include a number of spatially homogeneous Bianchi
models as special cases. The Bianchi models 
admit a three--dimensional group of isometries, $G_3$, that acts 
simply transitively on three--dimensional spacelike orbits \cite{rs}.  
The $G_3$ contains an Abelian subgroup $G_2$ for the types I--${\rm VII}_h$ 
and the LRS types VIII and IX \cite{tomita,carmeli}. 
Certain $G_2$ models may therefore be viewed as 
inhomogeneous generalizations of these Bianchi cosmologies.
In particular, for solutions of the form (\ref{psoln})--(\ref{c4}), 
 one has the following sub--classes of solutions:

\begin{enumerate}

\item The class with {\bf $\epsi=\epsii=m=\b=0$}  
corresponds to the homogeneous orthogonal Bianchi type I models 
containing a stiff perfect fluid \cite{jacobs}. 
They reduce to the Kasner solution in the vacuum 
limit $(\alpha =0)$ \cite{Kasner}. 

\item The class with {\bf $m=\b=0$, $\epsi=\epsii=1$} corresponds 
to an inhomogeneous generalisation of the Bianchi I models. 
Charach and Malin first 
considered solutions similar to these by 
imposing a three--torus topology on the spatial sections \cite{cm}. 
This effectively converts the integral over $l$ 
into an infinite sum, which also has the consequence of simplifying 
the solution of the remaining field equations for $f$. 
Adams {\em et al.}~\c{Adams1} further considered the vacuum case $(\phi=0)$. 

\item The class with {\bf $\epsi=\epsii=0$}, subject to the 
constraint $ \b^2-\al^2+m^2-k^2-3=0$, 
where  $k \equiv (-h)^{-1/2}$. In general, these 
models represent tilted stiff perfect fluid Bianchi type 
${\rm VI}_h$ cosmologies \cite{wim}, since the fluid velocity vector 
associated with the dilaton field 
is not orthogonal to the group orbits (surfaces of 
homogeneity). They reduce to the Bianchi type III and 
V models when $k^2 =1$ and $k=0$, respectively \cite{mn}. 
In the vacuum limit $(\alpha =\beta =0)$, 
the solution reduces to the Ellis--MacCallum type ${\rm VI}_h$ 
cosmology \cite{em}. 

\item The class with {\bf $\epsi=\epsii=0$} corresponds 
to the inhomogeneous generalisations of the Bianchi 
type III, V and VI$_h$ models first considered by
Wainwright {\em et al.} \cite{wim}. 
In general, however, these models suffer from spacelike 
singularities and their status as physical cosmologies 
is uncertain \cite{G2review1}. 

\end{enumerate}

An alternative way of considering inhomogeneous generalisations of the 
Bianchi type III, V and ${\rm VI}_h$ models is to specify 
$\epsilon_1 = \epsilon_2 =1$ and impose the constraint 
\be
\label{6constraint}
\beta^2-\al^2 +m^2-k^2-3=0.
\ee
Vacuum solutions of this type were considered by 
Adams et al.~\c{Adams2}, who concluded that the inhomogeneous 
structure of the initial cosmic singularity could evolve into 
gravitational waves propagating over a homogeneous background at 
late times. In view of the ambiguities associated with 
interpreting the Wainwright {\em et al.} \cite{wim} 
solutions in a cosmological context, 
our primary interest in the present paper is in this 
new class of models. We also remark that a subset  of
inhomogeneous Bianchi I models (item 2) is also included within this class 
since these latter models correspond  to the particular solution $\b=m=0$.

The remaining field equations
(\r{efe1}) and (\r{efe2}) for the longitudinal metric function $f$ 
may now be solved in principle 
by substituting in the derivatives of Eqs. (\r{psoln}) 
and (\r{phisoln}). 
However, solving Eq. (\r{efe1}) is non--trivial, because the 
right-hand-side contains integrals over $l$ that   
originate from the integral wave-train terms of 
Eqs. (\r{psoln}) and (\r{phisoln}). Unfortunately, 
these can not be expressed in a closed form and some 
simplifying assumptions must therefore 
be made in order to proceed analytically.  

In view of this, we consider two separate schemes. In the first, 
we restrict the analysis to a single mode, $l$, which could be viewed
as dominating over all the other modes. 
The choice of one mode allows the integrals over $l$ to be 
dropped\footnote{Particular solutions of the field equations that  
consist of superpositions of two or more modes may also be found by employing  
a discrete summation 
over $l$ in place of the integral 
in Eqs. (\r{psoln}) and (\r{phisoln}).}. In the second case, we assume the 
amplitudes of each of the modes in Eqs. (\ref{psoln}) and 
(\ref{phisoln}) are equal, i.e., we specify $C_l=C$, $D_l =D$, etc.

We first integrated Eq. (\r{efe1}) to deduce
an expression for $f$ containing an unknown function 
of integration $f_1(z)$. In each of the cases we considered, 
it was found that Eq. (\r{efe2}) was then trivially 
satisfied for constant $f_1(z)$. 
We now proceed to present the two 
classes of solutions, together with their homogeneous limits. 

\subsection{Homogeneous Solutions \l{Hom}}

The homogeneous 
limit of these solutions is determined  by specifying 
$\epsilon_1=\epsilon_2 = 0$ in Eqs. (\r{psoln}) and 
(\r{phisoln}) \cite{wim}. Eq. (\ref{efe1}) can then be integrated to 
yield the longitudinal component of the gravitational field 
in the metric (\ref{canmetric}): 
\bea
\l{fhom}
f_{hom}&=&C_1+\frac{1}{2} 
\L(\al^2+\b^2+ k^2+m^2-1\R)\ln(\xi) -\frac{1}{2} 
\L(\b^2+ m^2\R)\ln(z^2-\xi^2)\nn\\
&\,&
-\L(\al\b+km \R)\cosh^{-1}\L(\fr{z}{\xi}\R), 
\eea
where $C_1$ is an arbitrary constant of integration and the constraint 
equation 
(\ref{6constraint}) applies. In terms of the variables (\ref{emcoords}) 
this component is given by 
\be
\label{fhomt}
h_{hom} = C_1+\fr{1}{2}\L(\al^2+\b^2+ k^2+m^2-1\R)\ln \sinh (2t) +
(km +\alpha\beta ) \ln \tanh t
\ee
in Eq. (\ref{tzmetric}). The transverse component of the metric 
and the dilaton field respectively take 
the following forms
\be
\label{phomt}
p_{hom} =-2kZ+ k \ln\sinh (2t)+ m \ln\tanh t 
\ee
\be
\label{phihomt}
\phi_{hom} = -2\alpha Z +\alpha \ln \sinh (2t) + 
\beta \ln \tanh t.  
\ee 
The hypersurfaces $t={\rm constant}$ represent the 
surfaces of homogeneity. Since for $\alpha \ne 0$, the dilaton 
field depends on the spatial 
variable, $Z$, the fluid velocity as measured by $u_{\mu} 
= \phi_{, \mu}/(-\phi_{, \nu}\phi^{, \nu} )^{1/2}$ 
is not orthogonal to the group orbits. Thus,  
this solution may be interpreted as a tilted stiff fluid type ${\rm VI}_h$ 
solution. 

\subsection{Single-mode solutions\l{Auto}}

After restricting the analysis to a single mode, $l$, 
the integral of Eq. 
(\r{efe1}) is readily found. The resulting solution may
be expressed in terms of its homogeneous, gravitational wave 
and scalar wave components:
\be
\label{fsolution}
f= f_{hom}+f_{gw}+f_{sw}  ,
\ee
where the homogeneous component,  
$f_{hom}$,  is given by Eq. (\ref{fhom}), the gravitational wave 
component is given by 
\bea
\l{fgw}
f_{gw}&=&
+\fr{\xi^2}{4}\L(\L[c_1'J_0(l\xi)+c_2'N_0(l\xi)\R]^2
+\L[c_1'J_1(l\xi)+c_2'N_1(l\xi)\R]^2\R)
\nn\\&\,&
+\fr{l^2}{4}\xi^2\L(\L[c_1 J_0(l\xi)+c_2 N_0(l\xi)\R]^2+
\L[c_1 J_1(l\xi)+c_2 N_1(l\xi)\R]^2\R)
\nn\\&\,&
-\fr{l}{2}\xi\L[c_1 J_0(l\xi)+c_2 N_0(l\xi)\R]
\L[c_1 J_1(l\xi)+c_2 N_1(l\xi)\R]
+k\L(c_1 J_0(l\xi)+c_2 N_0(l\xi)\R)
\nn\\&\,&
-mlzc_1\int\fr{J_1(l\xi)}{\sqrt{z^2-\xi^2}}\,d\xi
-mlzc_2\int\fr{N_1(l\xi)}{\sqrt{z^2-\xi^2}}\,d\xi
\nn\\&\,&
-m c_1'\int\fr{\xi J_0(l\xi)}{\sqrt{z^2-\xi^2}}\,d\xi
-m c_2'\int\fr{\xi N_0(l\xi)}{\sqrt{z^2-\xi^2}}\,d\xi 
\eea
and the scalar-wave component $f_{sw}$ is determined 
by making the substitutions 
\be
\l{fswsubs}
k\to \al,\quad m\to \b,\quad c_1(z)\to c_3(z),\quad c_2(z)\to c_4(z) 
\ee
in Eq. (\ref{fgw}). The gravitational and scalar wave components
encode all the inhomogeneous contributions
to the longitudinal component of the gravitational field.
The integrals in the expressions for $f_{gw}$ 
and $f_{sw}$ can actually be performed, provided 
the integrand is first expressed as a series. The result is 
a series which reduces to a closed form expression in the 
asymptotic late-time limit.

\subsection{Equal-amplitude solutions \l{Demo}}

When all the modes in the integral wave--trains of Eqs. (\ref{psoln}) 
and (\ref{phisoln}) have equal weighting, we may employ the identities 
\be
\int_{0}^{\infty} J_0(l\xi)\,f(lz)\,dl = 
\L\{
\begin{array}{c l l}
\fr{1}{\sqrt{z^2-\xi^2}}, &\rm{if} & f(lz)=\sin(lz)\\
		       0, &\rm{if} & f(lz)=\cos(lz)
\end{array}\R.
\ee
\be
\int_{0}^{\infty} N_0(l\xi)\,f(lz)\,dl = 
\L\{
\begin{array}{c l l}
\fr{2}{\pi}\fr{\cosh^{-1}(z/\xi)}{\sqrt{z^2-\xi^2}}, &\rm{if} & f(lz)=\sin(lz)\\
\fr{-1}{\sqrt{z^2-\xi^2}}, &\rm{if} & f(lz)=\cos(lz)
\end{array}\R.
\ee
in order to integrate Eq. (\ref{efe1}). To avoid unnecessary complications, 
we further restrict our attention to 
the class of solutions where the contributions from the Neumann 
functions vanish $(c_2 =c_4 =0)$\footnote{The solution may also be found
for the case when the Neumann functions are included.}. 
The transverse component of the gravitational 
field and the dilaton are then given by 
\be
\label{democraticp}
p=k\ln(\xi)-m\cosh^{-1}\L(\fr{z}{\xi}\R)
+\fr{F}{\sqrt{z^2-\xi^2}}
\ee
\be
\label{democraticphi}
\phi=\alpha\ln(\xi)- \beta 
\cosh^{-1}\L(\fr{z}{\xi}\R)
+\fr{M}{\sqrt{z^2-\xi^2}}  ,
\ee
respectively, where $F$ and $M$ are arbitrary constants. 
The longitudinal component of the gravitational field, $f$, may  
now be determined by substituting Eqs. (\ref{democraticp}) and 
(\ref{democraticphi}) into Eqs. (\ref{efe1}) and (\r{efe2}) and integrating.
One finds that 
\be
\l{democraticf}
f=f_{hom}
+\fr{F^2+M^2}{4(z^2-\xi^2)^2}\xi^2
+\fr{mF+\b M}{z^2-\xi^2}z
+\fr{kF+\al M}{\sqrt{z^2-\xi^2}}.
\ee  
In terms of the variables defined in Eq. (\ref{emcoords}), 
it is given by
\be
\l{demoh}
h = h_{hom} + (kF+\alpha M) e^{2Z} + (mF + \beta M) e^{2Z} 
\cosh (2t) + \frac{1}{4} (F^2 +M^2) e^{4Z} \sinh^2 (2t)  ,
\ee
where $h_{hom}$ is given by Eq. (\ref{fhomt}). 
The transverse component and dilaton are given by 
\be
\l{demop}
p =-2kZ+ k \ln\sinh (2t)+ m \ln\tanh t + Fe^{2Z}
\ee
\be
\l{demophi}
\phi = -2\alpha Z + \alpha \ln\sinh (2t) + 
\beta \ln\tanh t + Me^{2Z}
\ee
when expressed in terms of these variables. 

\section{Asymptotic Behaviour}

\setcounter{equation}{0} 
\def\theequation{\thesection.\arabic{equation}}

In this section we consider some aspects of the asymptotic 
behaviour of the inhomogeneous string cosmologies derived above. 
The axion field in the 
dual solution (\ref{ds2-strings})--(\ref{ssigmavacuum})
tends to a constant value in the limits
$\phi \rightarrow \pm \infty$. It is therefore dynamically 
negligible in these limits. Eq. (\ref{sphivacuum}) implies that for 
all solutions generated by the ${\rm SL}(2,R)$ transformation 
(\ref{chi}), there exists a lower bound 
on the value of the dilaton field if $C$ and $D$ are non--zero. 
This implies the existence of a {\em lower} (non-vanishing) bound on the 
string coupling which, in the context of M--theory, in turn
implies the existence of a lower bound on
the radius of the eleventh dimension\footnote{We are assuming implicitly
that the extra six spatial dimensions are fixed.},
${\cal{R}}_{11}$, since ${\cal{R}}_{11}\propto e^{\phi /3}$ 
\cite{Mth,witten}.

In the class of dual 
solutions (\ref{tstringmetric}), it follows from Eq. (\ref{taxion}) 
that the two--form potential tends to a constant 
in the limits where $\Gamma \rightarrow 0$ and $\Gamma \rightarrow \infty$, 
i.e., when the volume of the transverse space measured 
in the string frame becomes vanishingly 
small or arbitrarily large. Thus, the two--form potential effectively 
decouples from the field equations in these limits. 
The limiting behaviour of the dilaton field in the dual 
solution follows from Eq. (\ref{tphi}). In particular, 
when $\Gamma \rightarrow 0$ and $d\ne 0$, the dilaton asymptotically 
tends to the form it has in the seed solution (up to a constant 
linear shift). The same conclusion 
applies for $\bar{\Gamma}$. In this limit, therefore, the dual 
cosmology asymptotes to the original 
seed, dilaton--vacuum solution. In the opposite
limit, where $\Gamma$ diverges, the dual cosmology 
asymptotes to the solution that is derived by 
specifying $d=0$ in Eq. (\ref{trho}). 

The above discussion applies to any class of 
string cosmologies derived with the symmetry transformations 
discussed in Section 2. To illustrate this further, we consider the 
asymptotic behaviour of the string cosmologies 
(\ref{tstringmetric}) derived 
from the equal--amplitude dilaton--vacuum, seed solution 
(\ref{democraticp})--(\ref{democraticf}). 

We recall that in determining the asymptotic behaviour of cosmological models
the choice of time gauge is important.
Since the models considered here may be viewed as
describing inhomogeneous waves propagating over 
homogeneous Bianchi backgrounds,
a reasonable measure of early and late times is provided by $t$ in
the coordinate chart $\{t,Z,x,y\}$.
As we shall see,  the asymptotic behaviour in 
the limits of small and large $t$ corresponds to either 
$\Gamma \rightarrow 0$
or $\infty$ in the solution (\ref{tstringmetric}). This 
implies that the class of string cosmologies (\ref{tstringmetric}) 
asymptotes 
between two dilaton--vacuum solutions, where the two--form is dynamically 
negligible. In effect, this field induces the transition between the 
two dilaton--vacuum limits.  A similar conclusion holds for the dual 
solution (\ref{ds2-strings}). 

In the limit, $t \rightarrow +\infty$, the relevant terms 
in the asymptotic forms 
of Eqs. (\ref{democraticp}), (\ref{democraticphi}) and 
(\ref{demoh}) are 
\bea
\label{largep}
p &\approx & 2k(t-Z)+F\,e^{2Z} 
\\
\label{largephi}
\phi &\approx &  2\al(t-Z)+M\,e^{2Z} 
\eea
\be
h \approx (\al^2+\b^2+k^2+m^2-1)t+\frac{1}{16}(F^2+M^2) e^{4(Z+t)} .
\ee
In terms of the coordinate pair $\{ \xi , z \}$, this implies that 
\bea
p &\approx & k\ln(\xi)+\fr{F}{\sqrt{z^2-\xi^2}} 
\\
\l{xiphi}
\phi &\approx &  \al\ln(\xi) +\fr{M}{\sqrt{z^2-\xi^2}} 
\eea
\be
\l{latedemofxiz}
f \approx  \fr{1}{2}
(\al^2+\b^2+k^2+m^2-1)\ln(\xi)+\fr{(F^2+M^2)\xi^2}{4(z^2-\xi^2)^2} .
\ee
We note that, in view of Eq. (\ref{emcoords}),
the singularity at $z=\xi$ only 
corresponds to  $(\xi,z) \rightarrow (\infty,\infty)$ and 
therefore cannot be reached in a finite $t$-time along {\em any} curve. 

In the limit $t \rightarrow 0$, the corresponding limiting forms 
of the metric components and dilaton field are 
\bea
\label{smallp}
p &\approx & -2kZ +(k+m) \ln t +Fe^{2Z} 
\\
\label{smallphi}
\phi &\approx & -2\alpha Z +(\alpha +\beta ) \ln t +Me^{2Z} 
\eea
\be
\label{smallf}
h \approx\fr{1}{2}\L[ \al^2 +\b^2 +k^2 +m^2 +2(km +\al\b) -1 \R]\ln(t) 
 +\left[ (k+m) F +(\alpha +\beta )M \right] e^{2Z} .
\ee
The limits of the determinant 
$\Gamma$ are deduced by substituting Eq. (\ref{emcoords}) 
and Eq. (\ref{largephi}) or (\ref{smallphi}) into Eq. (\ref{detvol}): 
\bea
\label{largeGamma}
\Gamma (t \rightarrow +\infty) &\approx& \exp 
\left[ 4(\alpha +1 )(t-Z) +2Me^{2Z} \right] 
\\
\label{smallGamma}
\Gamma (t \rightarrow 0 ) &\approx& 4\exp \left[ -4(1+\alpha )Z +2Me^{2Z}
+2(1+\alpha +\beta ) \ln t \right]   .
\eea
The behaviour of the dilaton field in the limit where 
the determinant diverges is most readily determined by substituting 
Eq. (\ref{detvol}) into Eq. (\ref{tphi}). We find that $\bar{\phi} 
\approx -\phi -2\ln \xi  -2\ln c$. This is interesting, because
for finite $Z$, Eq. (\ref{largephi}) 
implies that $\phi \propto \alpha \ln \xi$ 
for $t \rightarrow \infty$. For $\alpha \gg 1$, therefore, 
$\bar{\phi} \approx - \phi$ and, in this sense, the strongly 
coupled  limit of the dual solution may be viewed as the 
weakly coupled limit of the seed solution, and vice--versa. 

In taking the early or late time limits of an inhomogeneous
cosmological model, one has to take account of both
direction and time. Here, we have
inhomogeneity in the $Z$-direction only. 
The question that then arises is whether the $Z$-dependent terms in 
the above expressions dominate over the $t$--dependent ones. 
For any finite $Z$, the time--dependent terms eventually 
dominate as $t\rightarrow \{ 0, \infty \}$. 
A possible ambiguity in the limit arises, however, if we allow  
$Z(t)\to\infty$ sufficiently fast relative to the 
$t$--dependent terms. 

It is helpful to consider the simple set of straight lines 
${\cal V} = \{ Z=\rho t+\kappa~|\rho,\,\kappa\in {\cal R}\}$, as probes 
with which to study the asymptotic behaviour\fn{The 
precise form of these curves is not 
important, their utility derives from the fact that 
they may be employed to probe the three dynamically 
important cases $Z(t)\to \pm \infty,\, Z(t)=\hbox{finite}$.}. 
In this case, it follows from Eq. (\ref{largeGamma}) that 
\be
\l{Gammalimit}
\Gamma (t\to \infty) \approx\L\{
	\begin{array}{l c l}
	\exp[4(1+ \alpha )(t-Z)] &{\rm if}\quad &\rho\le 0 \\
	\exp[2M e^{2Z}\,]        &{\rm if}\quad &\rho >0
	\end{array}
	\R.  .
\ee
The sign of $\rho$ 
determines the term that dominates in Eq. (\ref{Gammalimit}),
with $\rho >0$ $(\rho<0)$  
corresponding to motion in the positive 
(negative) $Z$--direction while 
$\rho=0$ corresponds to moving along 
$Z={\rm constant}$ trajectories. 
Thus, for $\rho\le 0$, 
$\Gamma \to \infty$ for $\alpha > -1$, while $\Gamma \to \infty$ 
for $M>0$ and $\rho > 0$. On the other 
hand, the determinant becomes vanishingly small if 
$\alpha <-1$ and $\rho\le 0$ or if $M<0$ and $\rho >0$. 

In the early time limit, $Z\rightarrow \kappa={\rm constant}$ for the 
paths we are considering. Eq. (\ref{smallGamma}) then implies 
that $\Gamma \propto t^{2(1+\alpha +\beta )}$. 
It follows that $\Gamma \rightarrow 0$ if 
$\alpha +\beta >-1$ and $\Gamma \rightarrow \infty$ 
if $\alpha +\beta < -1$. 
The limiting forms of the dilaton and 
the transverse and longitudinal components of the metric 
are determined by allowing $Z \rightarrow \kappa$ in Eqs. 
(\ref{smallp})--(\ref{smallf}). Transforming to 
the synchronous frame by defining 
\begin{equation}
\tau \equiv \int^t dt' e^{h(t')/2} 
\end{equation}
then implies that, for both seed and dual solutions,  
the $G_2$ line-element (\r{g2line}) 
in either the Einstein or string frames qualitatively 
takes the form 
\be
\l{EarlyAutoline}
ds^2=-d\tau^2+A_1(\kappa)\tau^{a_1}dx^2+A_2(\kappa)\tau^{a_2}dy^2
      +A_3(\kappa )\tau^{a_3}dz^2,  
\ee
where the constants $a_i$ can be expressed in terms of  
$\{ k,m,\alpha , \beta \}$ and $A_i$ depend 
on $\kappa$. This solution represents 
an inhomogeneous generalization of 
the Kasner-Belinskii-Khalatnikov (KBK) solution~\c{BK},
in the sense that at each $Z$, parametrised here by the constant $\kappa$,
the universe describes a particular KBK solution. 

Thus, the determinant of the transverse 
space in the string frame vanishes or diverges in the early 
and late time limits. The precise behaviour depends 
on the constants that arise in the seed dilaton--vacuum 
solution. These constants are arbitrary modulo 
the constraint equation (\ref{6constraint}). In both limits, 
the two--form potential asymptotes to a constant value  and 
the dilaton field tends to its original form when 
$\Gamma \rightarrow 0$.

\section{Conclusion and Discussion}

\setcounter{equation}{0} 
\def\theequation{\thesection.\arabic{equation}}

In this paper we have employed the global symmetries of 
the string effective action to derive two families of inhomogeneous 
string cosmologies from a generalized Einstein--Rosen metric  
admitting an Abelian group of isometries, $G_2$. The solutions 
were parametrized in terms of the metric functions of the 
seed solution. Thus, the qualitative behaviour of these 
models can be determined directly from the asymptotic 
form of the original metric. 
Inhomogeneous generalizations of the Bianchi type ${\rm VI}_h$
string cosmologies containing a non--trivial dilaton and
two--form potential were derived. In general, these
fields induce inhomogeneities that may be
viewed as scalar and gravitational waves propagating over a
homogeneous background.

One of the main applications of Einstein-Rosen models arises because 
many of the non--perturbatively {\em exact} 
superstring backgrounds constructed from the gauged 
Wess--Zumino--Witten (WZW) models admit two 
abelian isometries\fn{See, for instance, Ref. 
\cite{tseytlin} for a review of the gauged WZW models.}. Hence, given 
our current knowledge of string cosmology and conformal field theory, 
the $G_2$ Einstein--Rosen cosmologies derived 
within the context of the low energy effective 
action represent a set of models that is closely related to many 
exact string solutions in four dimensions. 
In this sense, it is of primary importance to understand the
dynamical behaviour of these solutions. In particular, 
one question that arises is whether the $G_2$ solutions to 
the low energy string equations of motion
discussed in this work asymptote towards 
known WZW models in the early-- or late--time limits. 

Since our solutions are valid in the weak coupling regime, they 
should correctly represent the asymptotic states of perturbatively 
exact string cosmological models at future, or past, timelike infinity. 
On the other hand, in the strong coupling regime, i.e. near 
the big-bang singularity, one expects that higher--order 
corrections to the perturbative theory, as well as non--perturbative 
string effects, should become increasingly important. 
Hence, in this regime the qualitative behaviour of these solutions 
may deviate somewhat from that of solutions derived 
from the full M/string theory. However, there are reasons to believe 
that $G_2$ solutions should nevertheless provide a generic 
description of cosmological models 
in the vicinity of a singularity. A major incentive for 
this comes from the long standing conjecture of Belinski 
and Khalatnikov~\c{BK,BLK,BergerMay98}. This states that 
on the approach to the cosmological singularity, the 
generalized Einstein-Rosen metrics may play the role of the 
leading--order approximation to 
the {\em general} solution of conventional Einstein gravity. 
It is therefore important to study the behaviour of these 
models in the $t\to 0$ limit as well. 

In addition such solutions provide a useful framework within which to study
a number of other topics in early universe string cosmology. In particular, 
they serve as a theoretical setting for investigating 
the pre--big bang inflationary scenario \cite{pbb}. 
In the context of this scenario, one is usually interested 
in solutions where both the curvature and 
effective string coupling (the dilaton) diverge as $t\rightarrow 0$. 
So, from Eq. (\ref{smallphi}), it follows that one 
needs $\alpha +\beta < 0$ in order to satisfy the requirements 
for pre--big bang inflation in the 
string cosmologies generated from 
the equal--amplitude, dilaton--vacuum solutions. These solutions  
possess a Kasner-like early time limit
which under time reversal makes them compatible with 
pre-big bang inflation. However, in order 
for the inflation to end, this scenario requires a 
mechanism for inducing a graceful exit into the standard, post--big 
bang phase. At present, this problem is unresolved and it 
is possible that the inhomogeneous 
singularity may pose a further complication in addressing the exit problem.
It would be interesting to 
consider the homogenization of the universe within the 
context of these solutions. 

The initial conditions for the pre--big bang scenario are 
based on the assumption that both the curvature and coupling are 
sufficiently small enough to ensure that the universe is in 
the perturbative regime.
An important question in connection with this scenario
is the naturalness of these initial conditions \cite{clt}. 
An attempt has recently been made to address
this question by conjecturing that the past attractor of the inflationary 
solution is likely to be that of the Milne universe \cite{qh}.
If true,  this would go some way towards establishing 
the `naturalness' of the scenario. More general initial states 
were also recently considered \cite{qh1,tavakol}. 
The solutions discussed above possess a range of possible
late time limits, 
which in general are not Milne--like.
This suggests that this class of models 
is not generally compatible with the   
conjecture of Ref. \cite{qh}. 

Furthermore, there exists a lower 
bound on the string coupling for 
models generated by the ${\rm SL}(2,R)$ 
transformation (\ref{chi}) and this has implications for
the range of initial values that such a parameter can take. 
This in turn leads to an upper limit 
on the amount of inflation that can occur 
before higher--order effects become significant \cite{clt}. 

The solution generating techniques discussed in Section 
2 can be incorporated into more general algorithms. 
Although the models presented in this paper break 
spatial homogeneity along one direction, they still exhibit 
a certain degree of symmetry and it is important to 
develop further techniques that lead to more general 
solutions. Recently, by extending a 
previous method \cite{previous}, an algorithm was presented that 
generates inhomogeneous $G_1$ scalar field cosmologies 
exhibiting a single isometry from matter filled and vacuum $G_2$ backgrounds
\cite{lazkoz}. 
Such models break homogeneity in two spatial directions.
The discussion of Ref. \cite{lazkoz} was placed within the context of 
Einstein gravity and it would 
be interesting to adapt this algorithm to string cosmology. 
In principle, a family of inhomogeneous 
$G_1$ string cosmologies could then be generated from 
the $G_2$ solutions discussed in Section 2. 

The string effective action exhibits a further discrete
symmetry when there exits a $G_2$ isometry \cite{bakas}. 
This `mirror' symmetry 
interchanges the transverse metric degrees of freedom with 
the dilaton and axion fields and leads to a
new solution with a different spacetime interpretation \cite{bakas,lidsey}. 
When the axion field is trivial, the symmetry 
reduces to the simultaneous 
interchange $p \leftrightarrow \phi$ in Eqs. (\ref{efe1})--(\ref{efe4}). 
In particular, the 
negatively curved Friedman--Robertson--Walker string cosmology 
\cite{clw} may 
be generated in this way from a vacuum 
Bianchi type V model \cite{lidsey}. A third class of inhomogeneous 
string solutions with a non--trivial two--form potential 
may therefore be found by applying this discrete 
symmetry to the $G_2$ backgrounds derived in this work. 

The $G_2$ backgrounds we have discussed are parametrized by 
non--trivial fields from the NS--NS sector of the string effective 
action. This sector is common to all five perturbative string effective 
actions and the solutions may therefore be viewed as truncated solutions 
of both the heterotic and type II theories. The type IIB theory 
also has a non--trivial Ramond--Ramond (RR) sector, consisting of 
an additional axion field and a two--form potential 
\cite{GSW}. These fields
differ from those of the NS--NS sector in that they do not 
couple directly to the dilaton field in the effective action 
\cite{witten,bho}.  
There are further symmetry transformations that 
can be applied to generate a non--trivial RR sector from a
given NS--NS background \cite{copeland}. Our solutions therefore 
represent seeds for investigating the role of RR fields in inhomogeneous 
string cosmologies. To date, such fields have only been studied in an 
homogeneous setting.

\vspace{.3in}
\centerline{\bf Acknowledgments}
\vspace{.1in}
We thank W. B. Bonnor, M. A. H. MacCallum and M. A. Vazquez--Mozo for 
helpful comments and discussions. D. C. was 
supported by the Particle Physics and Astronomy 
Research Council (PPARC), A. F. was supported by Spanish 
Science Ministry Grant 172.310-0250/96, J.E.L. was supported by the Royal 
Society and R.T. benefited from PPARC UK Grant No. L39094. 
D.C. thanks the theoretical physics department at the University 
of Pa\'is Vasco for hospitality.

\end{document}